\pdfoutput=1
\documentclass[preprint,aps,showpacs,preprintnumbers,amsmath,amssymb]{revtex4}
\usepackage{graphicx}% Include figure files
\usepackage{dcolumn}% Align table columns on decimal point
\usepackage{bm}% bold math
\usepackage{color}
\begin{document}

\title{Electrical manipulation of the edge states in graphene and the effect on the
quantum Hall transport} 
\author{B. Ostahie$^{1,2}$, M. Ni\c t\u a$^{1}$ and A. Aldea$^{1,3}$}
\affiliation{$^1$ National Institute of Materials Physics, 77125 Bucharest-Magurele, 
Romania \\
$^2$ Faculty  of Physics, University of Bucharest, Romania\\
$^3$ Institute of Theoretical Physics, Cologne University, 50937
Cologne, Germany}
\date{\today}

\begin{abstract}
We investigate  new properties of the Dirac electrons in the finite graphene 
sample under perpendicular magnetic field that emerge  when  an in-plane 
electric bias is also  applied.
The numerical analysis of the Hofstadter spectrum and of the edge-type 
wave functions evidentiate the presence of  {\it shortcut edge states} 
that appear under the influence of the electric field. 
The states are characterized by a specific spatial distribution, 
which follows only partially the perimeter, and exhibit ridges  
that connect   opposite sides of the graphene plaquette. 
Two kinds of such states have been found in different regions of the spectrum, 
their particular spatial localization being shown  along with the diamagnetic 
moments that reveal their chirality.

By simulating a four-lead  Hall device, we investigate the transport properties
and observe new, unconventional plateaus of the integer quantum Hall effect,
which are associated  with the presence of the shortcut edge states.
The contributions of the  novel states to the conductance matrix 
that  determine the new transport properties are  shown. 
The shortcut edge states resulting from the splitting of the n=0 Landau level 
represent a special case, giving rise to non-trivial transverse and 
longitudinal resistance.
\end{abstract}
\maketitle

\section{Introduction}
The spectral and transport properties of graphene, including the topological 
aspects, were studied in the presence of the spin-orbit coupling \cite{Kane}, 
of the external magnetic field \cite{Novoselov, Kim, Goerbig}, 
and also considering  
different geometries, like the ribbon \cite{Peres, Nakada, Brey} or 
finite plaquette \cite{Igor,Onipko,Kramer, Ostahie}. 
In  geometrically confined systems, the edge states, of either chiral or  helical  
origin, play an  essential role acting  as  charge and spin  channels for 
the integer and spin quantum Hall transport at a given  Fermi energy. 

In this paper, we introduce as a supplementary ingredient an in-plane electric bias 
that allows for the  manipulation of the conducting channels, with immediate 
consequences on the quantum Hall effect. 
The role of the electric bias is to act on  the spatial position of the channels. 
Then, having in mind a many-terminal Hall device, it is obvious that the
migration of the channels with the electric field affects the electron 
transmittance between different leads, fact that suffices to change the 
quantum Hall plateaus.
The modified edge states, responsible for this effect, will be called {\it shortcut
edge states} for reasons that are obvious in Fig.7. 

The possibility  to  electrically manipulate the edge states generated by  
the magnetic field was  advanced in \cite{Aldea} for the confined  2D electron 
gas using a tight-binding approach. We remind that the 2D gas exhibits only 
conventional Landau bands, which depend linearly on the magnetic field. 
However, the study of graphene  looks  especially  promising
due to the specific aspects as the relativistic range of the  energy spectrum
(where the Dirac-Landau bands show the square root dependence on the magnetic field),
and the presence of the flat (independent of the magnetic field) n=0 Landau level 
at  zero-energy.

The numerical calculation of the spectral properties of the finite plaquette 
subjected to crossed electric and magnetic fields, corroborated by the calculation 
of the electron transmittance through the  Hall device (obtained by attaching four 
leads), indicate the presence of what we call shortcut edge states. Their distribution 
on the plaquette is such that the wave function is localized mainly along the edges,
but shows also a shortcut in the middle, the position of which being perpendicular 
on and controlled by the electric field. We have identified two kinds of 
such states, some being spread among the bulk states of the  relativistic 
Landau bands, and others resulting from the electrically induced degeneracy 
lifting of the n=0 Landau level.

The zero-energy Landau level attracted much interest, its splitting being obtained 
in different ways: by using an external magnetic field in the quantum
extreme limit \cite{Levitov, Fertig}, the internal magnetic field in the 
magnetic topological insulators \cite{Zhang, Zhang1} or by disorder \cite{Roche}. 
In all situation, the transverse conductance shows plateaus
near $E=0$, however, the longitudinal one is a problem under debate,
its behavior being either dissipative or showing sometimes a tendency towards 
non-dissipative character.
The n=0 LL is studied  intensively also in the context of the quantum Hall
ferromagnetism driven by the exchange interaction (see \cite{Kharitonov} 
and the references therein).

In this paper, the degeneracy lifting of the zero-energy Landau level is ensured
by the electric bias  applied in the plane of the graphene plaquette.
The result is a  Wannier-Stark ladder  composed of a sequence of  shortcut 
edge states with  alternating chirality (meaning that they carry opposite 
currents and show opposite diamagnetic moments).
Due to this property, one may assume the presence in the transverse (Hall) 
resistance of a quantum plateau  $R_H=0$ in the energy range  about $E=0$, and, 
indeed, this guess is confirmed by the  Landauer-B\"{u}ttiker -type calculation. 
In what concerns the longitudinal resistance, we find that $R_L$ may show 
dissipative or non-dissipative character, as depending on the configuration of the 
current and voltage terminals attached to the graphene plaquette.

We mention that, in order to  detect the formation mechanism of the shortcut 
edge states in the plaquette geometry, it was very useful to study first 
the effect of the electric field on the edge states in the zig-zag graphene ribbon. 
The analysis is presented in the next section.

Although our attention is paid to the novel edge states in crossed electric 
and magnetic fields  and to their influence on the transport properties,  
we  remark that  the effect of the external electric field in graphene
was studied also in some other contexts. 
For instance, one may ask how robust should the Landau spectrum be against the
applied electric field. The problem was addressed in Ref.\cite{Baskaran,Castro}, 
where it was  proved analytically that, in the  low energy range of the graphene 
spectrum the interlevel distance decreases with the electric field, and, eventually, 
for a critical value of the ratio $\beta$ between the electric 
field 
$\mathcal{E}$ and the magnetic field  $B$,
all Dirac-Landau levels collapse ($\beta=\mathcal{E}/v_F B$, 
$v_F$= Fermi velocity).
Another aspect, discussed in Ref.\cite{Kelardeh}, concerns the specific 
properties of the Wannier-Stark states in graphene under strong electric field. 
It was shown that the mixing between the  conduction and valence bands, 
induced by the electric field near the Dirac points, gives rise to an energy spectrum 
characterized by  anticrossing points when the electric field is varied.

Our paper is organized as follows: 
in section II, we calculate the energy spectrum and observe the specific behavior 
of the novel edge states that appear under the effect of the electric bias.
This is done  both for the graphene zig-zag ribbon and the plaquette geometry.
In section III we prove the  effect of the shortcut edge states on the 
electron conductance matrix, and the unusual aspect of the quantum Hall effect 
which shows novel plateaus.
These aspects are approached both heuristically and numerically in the frame of 
the Landauer-B\"{u}ttiker formalism. 
The conclusions can be found in the last section. 

\section{Spectral properties of the zig-zag nanoribbon and  finite graphene 
lattice  in the tight-binding model}
In this section, our goal is to bring out particular aspects of 
spectral properties of \textit{finite} honeycomb plaquette in the simultaneous 
presence of a strong perpendicular magnetic field and an in-plane electric field. 
The attention will be focused on the novel chiral edge states that appear 
under applied bias in the relativistic range of the graphene energy spectrum. 
The understanding of their specific arrangement on the plaquette 
imposed by the electric field is, however, more accessible if we discuss 
first the same problem in the ribbon geometry. 

%Fig1
\begin{figure}[!ht]
\centering
\includegraphics{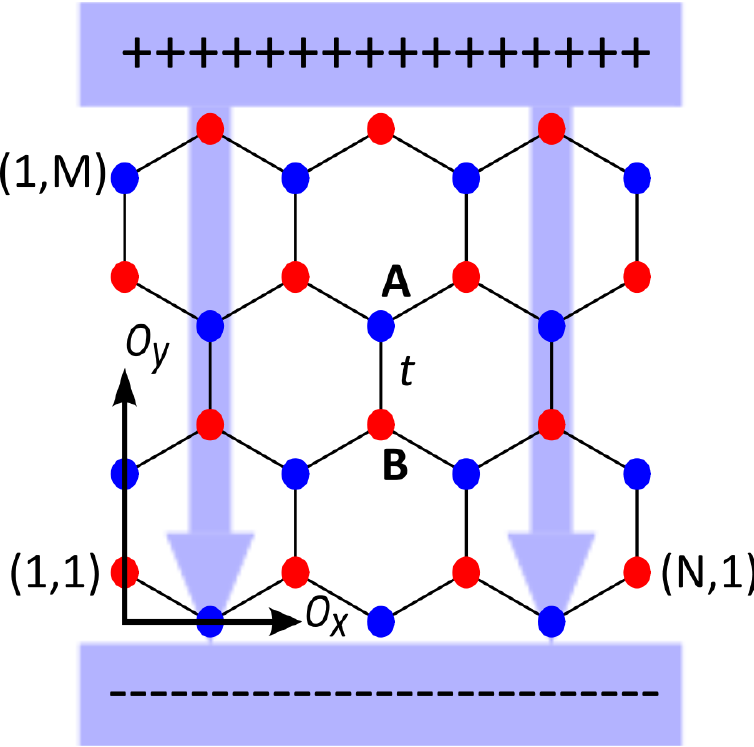}
\caption{(Color online) A piece of honeycomb lattice with two type of edges: 
zig-zag (along the  $\textbf{O}_{x}$ direction) and armchair 
(along the $\textbf{O}_{y}$ direction). 
The blue points belong to the sublattice \textbf{\textit{A}} and the 
red points to the sublattice \textbf{\textit{B}}, \textbf{\textit{t}} 
is the hopping amplitude connecting the nearest neighbor lattice points; 
the static electric field (blue arrows) is applied parallel  
to $\textbf{O}_{y}$ axis. The number of lattice sites is 
$N \times M$, were $N$ is odd and $M$ is even, in our sketch $7 \times 4$.}
\end{figure}

Since the honeycomb lattice arises from the  overlapping of two triangular lattices 
\textbf{\textit{A}} and \textbf{\textit{B}}, 
the tight-binding Hamiltonian can be written in terms of creation and annihilation 
operators $a_{n,m}^{\dagger}, a_{n,m}$ and  $b_{n,m}^{\dagger}, b_{n,m}$, 
which act on  the sites of the two sublattices, correspondingly. 
The counting of the atoms can be done in different ways. Here,
in accordance  with Fig.1, the index $n$ counts the atoms along the horizontal 
zig-zag chains, and $m$ is  the chain index ($n\in[1,N]$, $m\in[1,M]$). 
It turns out that for the \textbf{\textit{A}}-sites of the blue-sublattice 
one has $n+m=odd$, while for the red sublattice $n+m=even$. 
In the presence of a perpendicular magnetic field, the hopping integral $t$ 
(connecting the two sublattices) acquires a Peierls phase which can be calculated 
by integrating the vector potential along the A-B bonds. 
Then, the spinless tight-binding Hamiltonian that describes the $\pi$-electrons 
in the graphene lattice in  perpendicular magnetic field has the form:
\begin{eqnarray}
H=\sum_{\substack{n,m\\n+m=odd}}{\epsilon^{a}a^{\dag}_{n,m}a_{n,m}}
+\sum_{\substack{n,m\\n+m=even}}{\epsilon^{b}b^{\dag}_{n,m}b_{n,m}}
~~~~~~~~~~~~~~~~~~~~~~\nonumber \\ 
+t\sum_{n,m}\big({e^{\pi i\phi (m-\frac{5}{6})}a^{\dag}_{n,m}b_{n-1,m}
+e^{-\pi i\phi(m-\frac{5}{6})}a^{\dag}_{n,m}b_{n+1,m}}\big)
~~~~~~~~~~~~\nonumber \\
+t\sum_{n=odd}\sum_{m=even}{a^{\dag}_{n,m}b_{n,m-1}}+
t\sum_{n=even}\sum_{m=odd}{b^{\dag}_{n,m}a_{n,m+1}}+H.c. ,
\end{eqnarray}
where  $\phi$ is the magnetic flux through the hexagonal cell 
measured in flux quantum units $h/e$, and the vector potential was chosen  
as $\vec{A}=(-By,0,0)$. The first two terms in Eq.(1) represent the atomic 
contributions, 
the next two terms  describe the hopping along the zig-zag chains \cite{Note-phi},
while the last ones describe the hopping between the neighboring chains.

The in-plane electric field is introduced along the  $O_y$ direction, 
and it is simulated by replacing in Eq.(1) the on-site energies $\epsilon^{a,b}$  
with  $\epsilon^{a,b}_{n,m}=e\mathcal{E}(y_{n,m}-L_{y}/2)$, 
where $\mathcal{E}$ is the  electric field, $y_{n,m}$ is the site coordinate, 
and $L_{y}$ is the length of the honeycomb lattice along 
the $O_y$ direction ($L_y=(3M-2)a/2$, where $a$ is the hexagon 
side length). 
 
The spectral properties of the Hamiltonian depend on  the boundary conditions 
imposed to the wave function, and the most studied case is that one of the 
graphene ribbon, meaning that periodic conditions are imposed
along one direction only. In the next subsection, we shall discuss first  
the effect of the electric field on the edge states in the zig-zag ribbon, 
and then  extend the analysis to the finite plaquette, obtained by imposing 
vanishing boundary conditions all around the graphene lattice. 
Different behaviors of the edge states can be identified in three regions 
of the Hofstadter spectrum. At the extremities of the spectrum, 
in the conventional Landau range, the modifications of the edge states 
are similar to those discussed already in \cite{Aldea} for the case of the 
confined 2D electron gas.
Thus, in this paper, we concentrate on the relativistic Landau range, 
and also on the special case of the n=0 Landau level placed 
at the energy $E=0$. These two ranges  exhibit the specific behavior of the 
Dirac electrons in graphene.

\subsection{Zig-zag nanoribbon: spectral properties in crossed 
magnetic and electric fields}
In the ribbon case, the momentum $k_{x}$  is a good quantum number, and 
the energy spectrum will be obtained by diagonalizing a $2M \times 2M$ matrix
for any momentum $k_{x}$. 
The resulting eigenvalues as a function of $k_{x}$ are shown in Fig.2a 
in the absence of the electric field and in Fig.2b for a nonvanishing 
electric field. 
%Fig2
\begin{figure}[!ht]
\centering{
\includegraphics[scale=1]{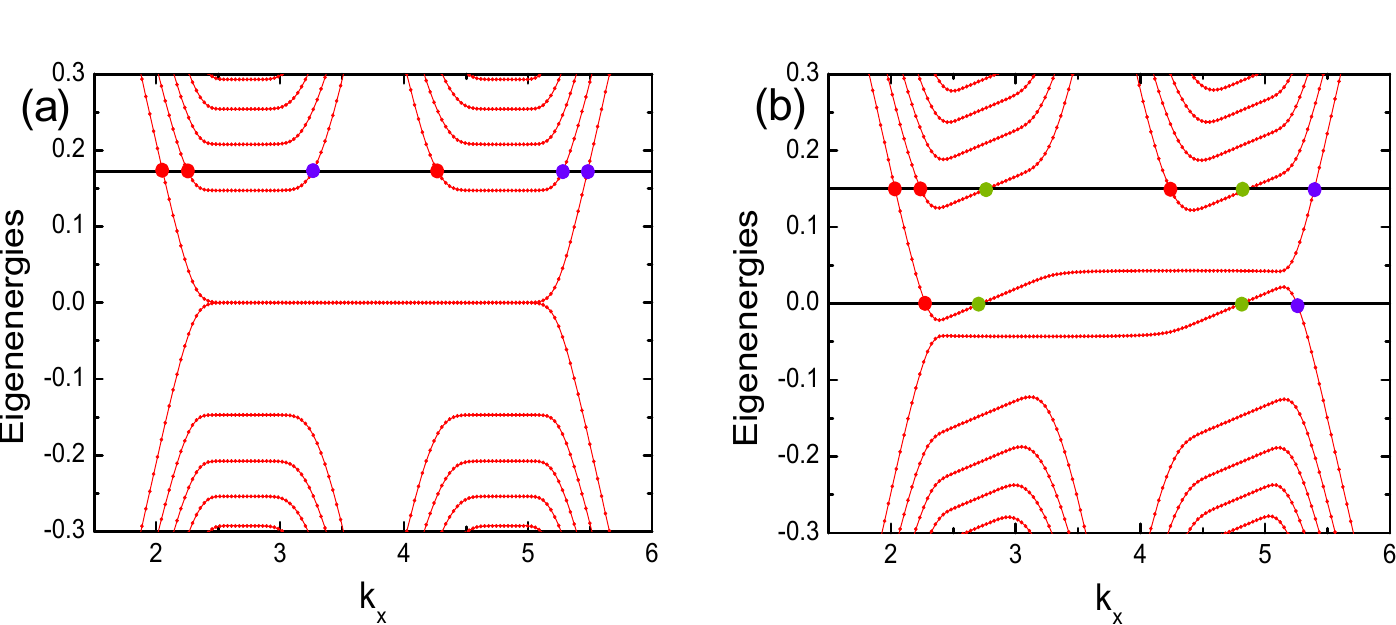}
\centering
\caption{(Color online) The energy spectrum of a graphene zig-zag nanoribbon: 
(a) in the presence of perpendicular magnetic field and (b) in the presence of both
magnetic and in-plane electric fields. The black lines correspond to the Fermi level
and the red, blue and green dots highlight the quantum states whose corresponding
conducting channels are represented in Fig.3. The theoretical simulation is performed
at the following parameters: the number of sites $400 \times 100$,
the magnetic flux $\phi/\phi_0=1/500$, the electric bias 
$e\mathcal{E}L_y/t=0.1$.
(These values correspond to the width $L_y=
10 l_B= 21 nm$, magnetic field $B= 158 T$, electric field 
$\mathcal{E}=1.3\times 10^7 V/m$, and a parameter $\beta=0.09$; 
experimentally accessible values are suggested below by the scaling 
law Eq.(5).)}}
\end{figure}

For the already studied case $\mathcal{E}=0$ \cite{Neto}, the energy spectrum
shows electron-hole symmetry with  degenerate flat Landau  bands,
whose wave functions are located in the middle of the stripe,
and dispersive states along  the edges (see Fig.2a).
Obviously, the non-zero velocity  
$v(k_x)=\frac{1}{\hbar}dE(k_x)/dk_x$ and its sign
indicate the presence of currents flowing along the edges in opposite directions.
For instance, choosing the Fermi level between $n=1$ and $n=2$ Landau levels, 
we notice 3 channels with  negative velocity (marked with red dots) and 
other 3 channels (marked in blue) flowing oppositely on the other side 
of the ribbon. The picture in the real space is illustrated in Fig.3a. 
%Fig3
\begin{figure}[!ht]
\centering{
\includegraphics[scale=0.9]{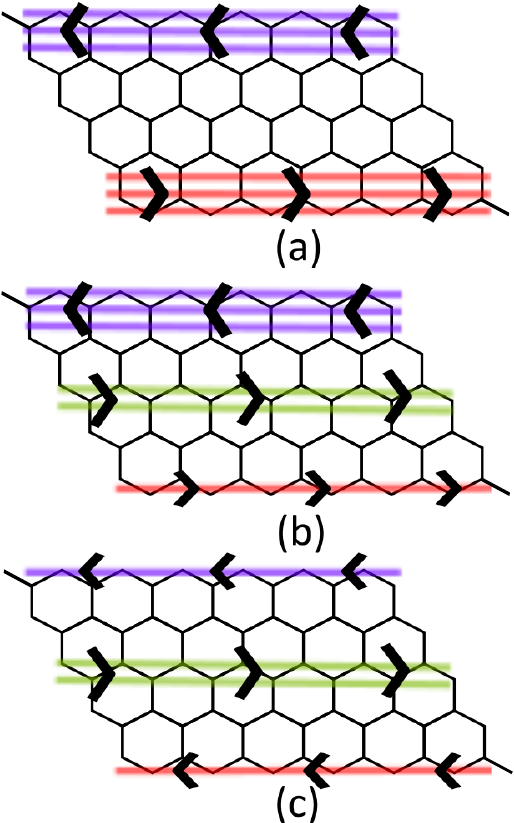}}
\centering
\caption{(Color online) Schematic representation of conducting channels in
the zig-zag nanoribbon geometry corresponding to the three situations
discussed in Fig.2. (a) At $\mathcal{E}=0$ there are three
edge channels on each side  of the ribbon, running in opposite directions.
(b) At $\mathcal{E}\ne 0$  and $E_F=0.15$, due to the tilt of the 
former flat band, two channels are pushed in the middle of the stripe.
(c) The special case   $\mathcal{E}\ne 0$  and $E_F=0$ shows only four
channels: unusually, the edge channels run in the same direction, while the 
two bulk channels run oppositely.}
\end{figure}

The additional electric field induces qualitatively new features. 
The former flat Landau bands get a tilt, which indicates the degeneracy 
lifting and a finite velocity $\frac{1}{\hbar}dE/dk_x$ 
(positive in Fig.2b for the chosen direction of the electric field). 
One may observe that  a partial degeneracy lifting occurs also to
the band $n=0$, suggesting new effects  around the energy $E=0$. 
However, the most significant observation at $\mathcal{E} \ne 0$ is that, 
because of the tilt of the spectrum, the number of edge channels 
on the two sides of the zig-zag ribbon becomes different, and
current carrying channels  appear also in the middle of the stripe at the 
expenses of the edge channels. 
The number and position of the channels can be observed by counting  
the intersections of the Fermi level with the spectrum branches in Fig.2b. 
Different situations may occur depending on the position of the Fermi level  
in the spectrum, and two specific cases can be highlighted:\\ 
i) at $E_f=0.15$ we notice three edge channels on the left 
(stemming from n=0 and n=1 Landau levels, and marked with red dots), 
one edge channel on the right (stemming from n=0, marked in blue, and running 
in opposite direction), but also two channels (marked with green dots) 
coming from the inclination of the former flat band. 
The numerical calculation of the eigenvectors reveals that the 'green' channels 
are located in the  middle of the stripe as shown in Fig.3b.\\
ii) at $E_f=0$ a very interesting case occurs.  Fig.2b puts into evidence 
two  edge states (marked in red and blue) which show this time the same 
(negative) derivative, i.e., the same direction of the current, 
while the 'green' doublet in the middle shows the opposite sign. 
The distribution of the channels in the real space turns out to be unusual 
in this case as shown in Fig.3c.

The three situations discussed for the  ribbon geometry show that
the location of the channels along the $O_y$  axis depends on the presence/absence 
of the electric field and on the position of the Fermi level.
We are left now with the question how the channels will  be distributed 
in the actual case of a finite rectangular plaquette. 

The energy spectrum of the graphene ribbon in crossed electric and 
magnetic fields apparently shows in Fig.2b the  symmetry 
$E(k^0_x+k_x)=-E(k^0_x-k_x)$ around a point $k^0_x$, which was also noticed  in 
\cite{Roslyak} in the frame of the  continuum model. 
In Appendix we prove the existence and calculate the value of $k^0_x$. 
The proof indicates that the spectrum symmetry relies on the
specific inversion symmetry of the honeycomb lattice that moves the
atoms A in atoms B and vice-versa. The position of  $k^0_x$ is defined
modulo $\pi$ and depends on the magnetic field $B$ and the ribbon width $M$.

We continue the discussion on the graphene ribbon with  considerations 
concerning  finite size aspects and scaling properties of the energy spectrum.
One knows that the two-dimensional electron gas  in a perpendicular magnetic 
field shows the degenerate Landau spectrum and quantum transport properties 
if the linear dimension $L$ is much larger than the magnetic length 
$l_B=\sqrt{\hbar/e B}$ ($L>>l_B$). In other words, at a given $B$, 
the system enters the 
Landau quantization regime
only if $L$ is sufficiently large. For the graphene ribbon, the gradual formation of 
the Dirac-Landau spectrum with increasing width  was noticed already in
\cite{Huang}. We return to  this problem in Fig.4a where we keep the magnetic flux 
fixed at $\phi/\phi_0=0.0005$ (corresponding to $B=39.5 T$) and observe the formation 
of the degenerate Dirac-Landau levels when the width $L_y$ increases. 
The only level that is always present, no matter the ribbon width, is the level $n=0$. 
In what concerns the others $n=1,2,...$, one can see that, for instance at M=100 
(when  the ribbon width $L_y=(\frac{3M}{2}-1)a= 21nm = 5.2 l_B$),  
the specific degenerate levels do not manifest themselves (the red  curve). 
However, such states appear at the  larger widths corresponding to M=200 and 400 
($L_y=42.4 nm\approx 10.5 l_B$ and $Ly=85 nm \approx 21 l_B$, respectively).

We make now  a step further proving  the finite size scaling behavior  of the 
spectrum in the relativistic domain. To this aim, one calculates the energy spectrum 
for different widths and magnetic fields and finds a scaling function 
that allows the superposition of all curves into a single one as in  Fig.4b. 
One makes first the scaling hypothesis that the
relativistic Landau eigenvalues $E_n$ are homogeneous functions of $B$ and $L_y$ :
\begin{equation}
\lambda E_n(B,L_y) = E_n(\lambda^{s_1}B, \lambda^{s_2} L_y)
\end{equation}
The parameter $\lambda$ is arbitrary and, with the choice  $\lambda=B^{-1/s_{1}}$,
one gets
\begin{equation}
B^{-1/s_{1}} E_n(B,L_y) = E_n(1,B^{-s_{2}/s_{1}} L_y).
\end{equation}
The scaling behavior shown in Fig.4b occurs for $s_1=2$ and $s_2=-2$, resulting
the following scaling law:
\begin{equation}
E_n(B,L_y) = \sqrt{B}f(B L_y).
\end{equation}
In Fig.4c we detected also the scaling  behavior of the low energy spectrum in the 
presence of both magnetic and electric fields. 
In this case, repeating the above arguments, the resulting scaling law is the following: 
\begin{equation}
E_n(B,\mathcal{E},L_y) = \sqrt{B}\tilde f(B L_y, \mathcal{E} L_y^{3/2}).
\end{equation}
%Fig4
\begin{figure}[htb]
\hskip-2.cm
\includegraphics[angle=-0,width=0.6\textwidth]{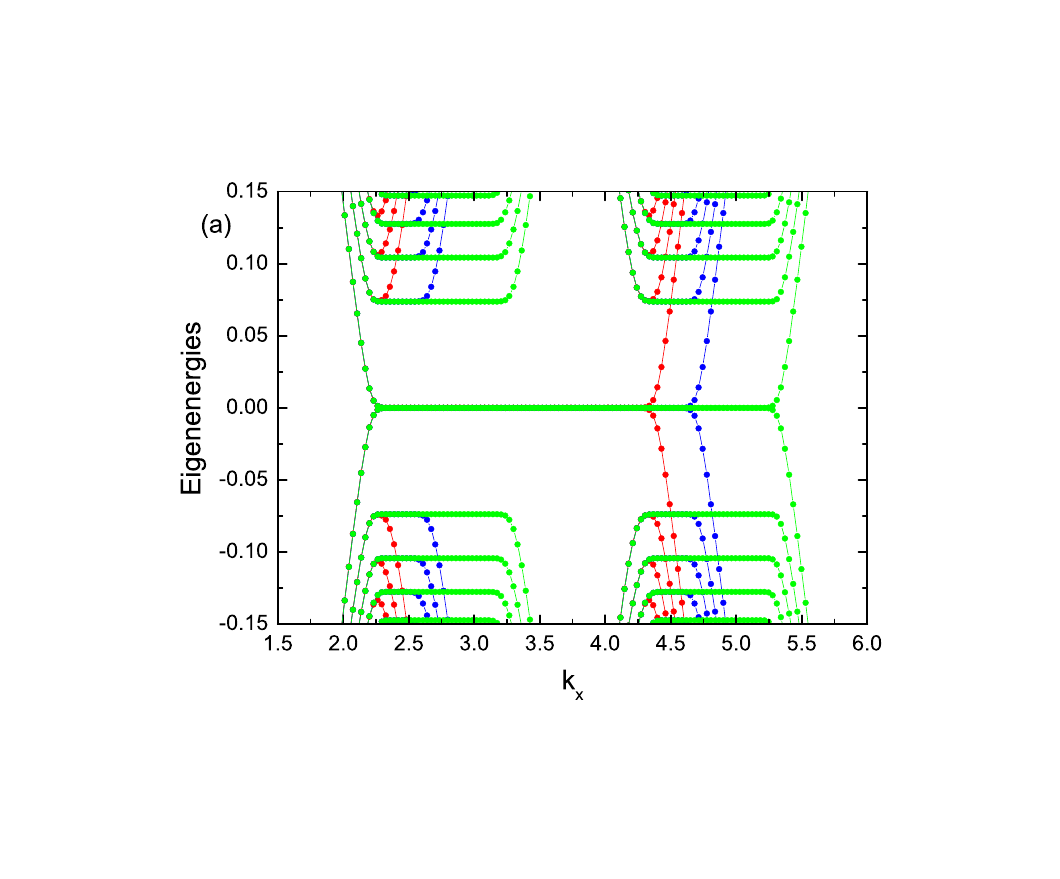}
\hskip-1.8cm
\includegraphics[angle=-0,width=0.6\textwidth]{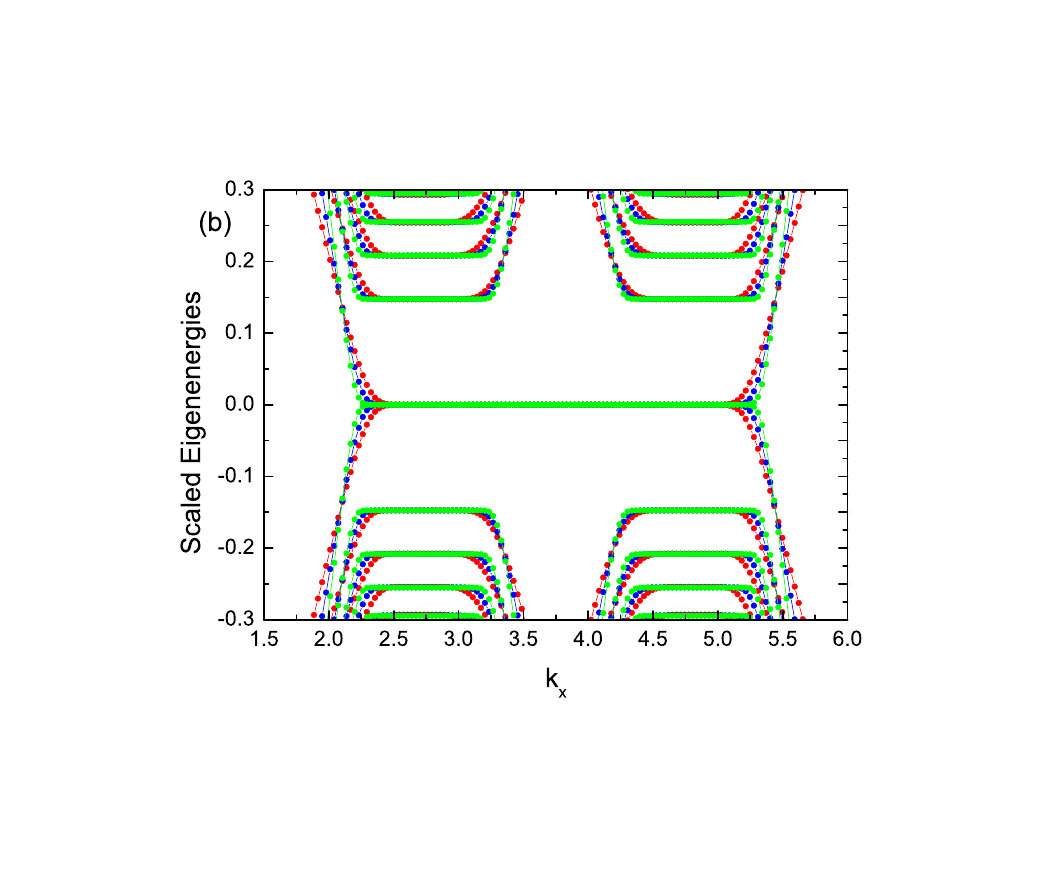}
\vskip-3cm
\hskip-2.7cm
\includegraphics[angle=-0,width=0.6\textwidth]{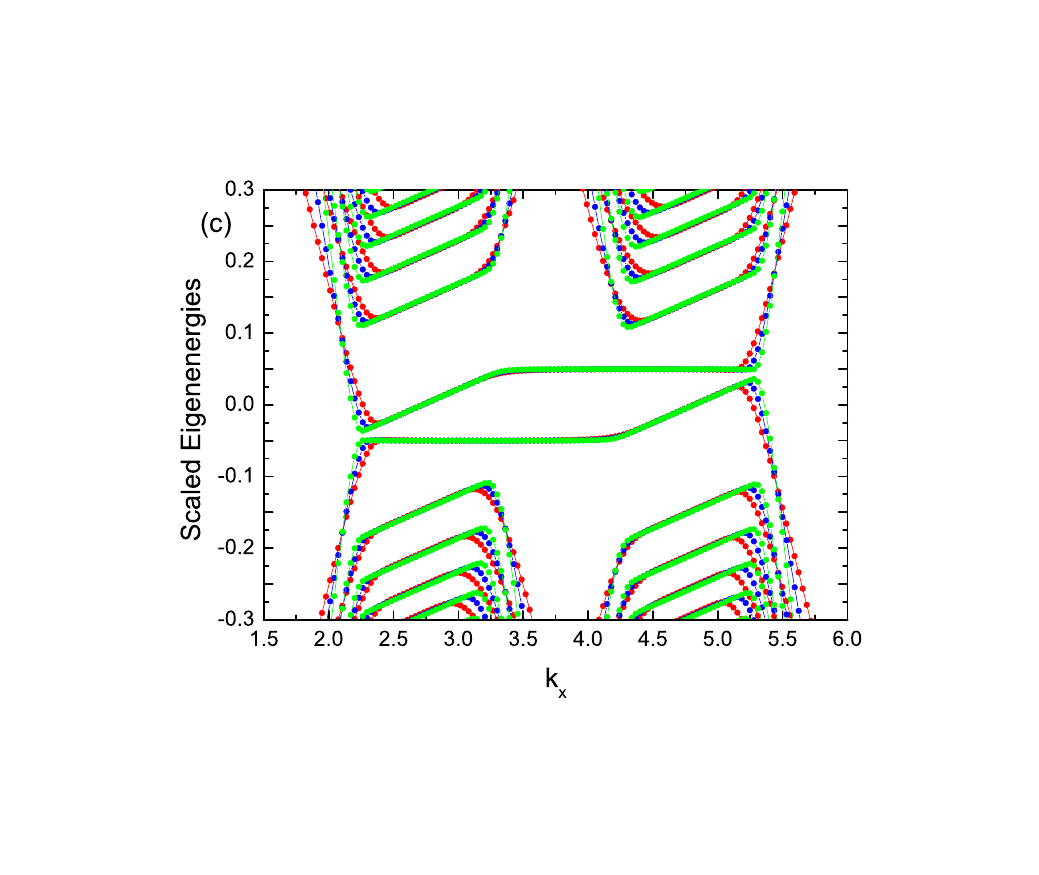}
\vskip-2cm
\caption{(Color online) 
(a) The gradual formation of the Dirac-Landau levels with increasing width of the
graphene ribbon at $\phi/\phi_0=0.0005~ (B=39.5 T)$: $M=100$ (red), $M=200$ (blue), $M=400$ (green). 
(b) The  eigenenergies scaled according to  Eq.(4); the three superimposed curves correspond to 
%according to Eq.(4) of the energy spectrum  in the presence of the magnetic field.
$M=100,\phi/\phi_0=0.002~(B=158 T)$ (red), $M=200,\phi/\phi_0=0.001~(B=79 T)$ (blue),
and $M=400,\phi/\phi_0=0.0005~ (B=39.5 T)$ (green). 
(c) The scaled eigenenergies according to Eq.(5) when both magnetic and
electric fields are present; the sequence of curves is the same as in (b), the electric bias 
being $e\mathcal{E}L_y/t=0.1(\mathcal{E}=1.3\times10^7V/m), 
e\mathcal{E}L_y/t=0.1/\sqrt{2} (\mathcal{E}=4.5\times10^6V/m)$, 
and $e\mathcal{E}L_y/t=0.05(\mathcal{E}=1.6\times10^6V/m)$, correspondingly.}
\end{figure}

The finite size scaling, besides being intrinsically interesting, helps to overcome
technical problems  met in the  numerical simulations of the physical effects.
Since the computer simulation of finite but large systems pretends much memory 
and running time, one has to consider smaller systems subjected however 
to external (magnetic, electric)
fields higher than available experimentally. Then, the scaling law says that the same 
behavior is expected at smaller fields but at larger size. For instance, the 
data in Fig.2b are obtained at $B= 158 T$ and $\mathcal{E}=1.3\times 10^7 V/m$ for a width 
of $L_y=21 nm$, but the scaling law Eq.(5) shows that the same value of the 
scaling function would be obtained at $B= 15 T$ and
$\mathcal{E}=4\times 10^5 V/m$, if the width were $L_y=221 nm$.

We note that  the analysis of the finite size scaling for the
graphene plaquette is technically  more difficult than for the ribbon, 
and it will not be addressed here.

\subsection{Shortcut edge states in finite plaquette geometry}
The finite size plaquette can be obtained from the ribbon  by  imposing 
vanishing boundary conditions also along the $O_x$ direction.
The resulting structure we shall consider will be like in Fig.1, the
rectangular geometry showing  two zig-zag and
two arm-chair boundaries.
Then one can assume that the channels, described above for the ribbon case, 
will give rise to closed circuits in the confined system. 
The main question is how the channels located in the middle of the stripe 
(shown in Figs.3b,c) will get closed in the plaquette geometry. 
Intuitively, they should generate  channels that touch all the four edges, 
but also channels that  get closed through the middle of the plaquette.
However, the  analysis performed in the previous subsection cannot 
be adopted now in a straightforward manner since the momentum $k_x$ is no 
more a good quantum number.
So, we shall discuss the effect of the applied electric field  in terms 
of the changes induced to the Hofstadter energy spectrum, to the  bulk and edge 
quantum states, and to the  diamagnetic response of the system.  

%Fig5
\begin{figure}[!ht]
\centering{
\includegraphics[scale=0.75]{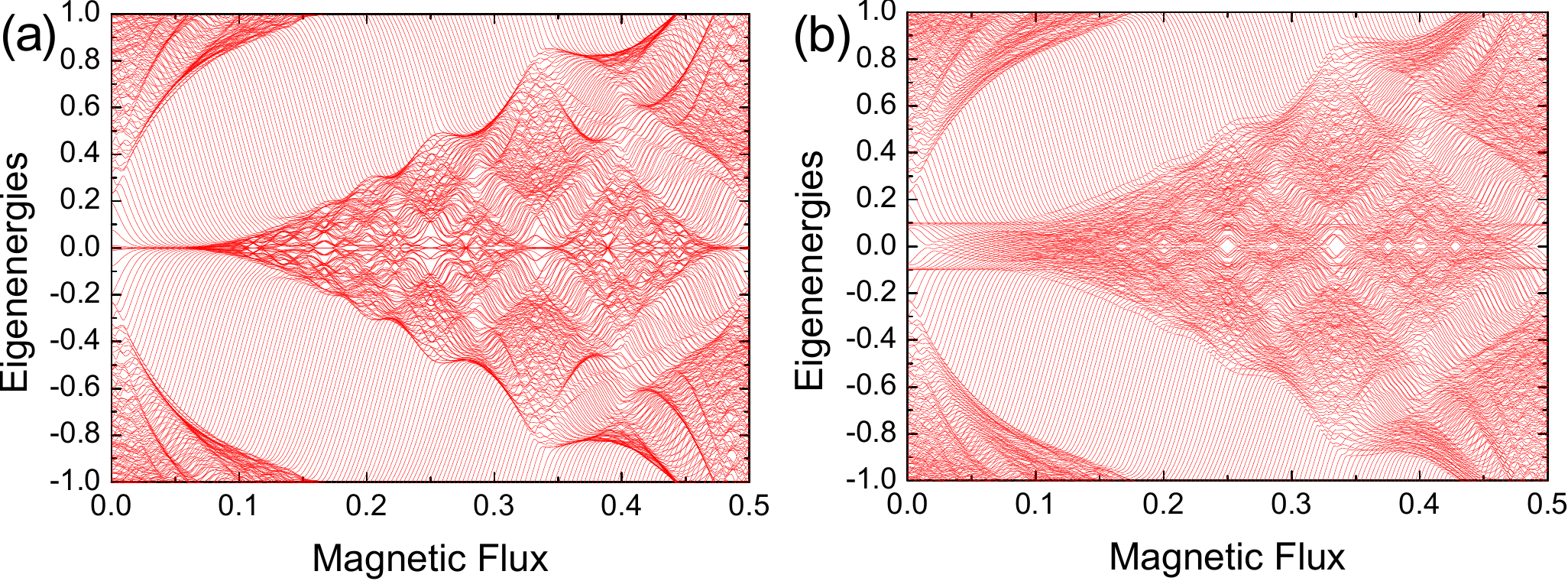}}
\centering
\caption{ (a) The relativistic range of the Hofstadter spectrum for a graphene 
plaquette in the absence of the electric field. Edge states fill the gaps
as an effect of the finite size. (b) The in-plane electric field lifts
the quasi-degeneracy of the Landau bands; the most visible effect
concerns the n=0 band in the middle of the spectrum. The number of lattice sites 
is $35\times20$, and the electric bias is $e\mathcal{E}L_y/t=0.2$.
(The energy is measured in units of hopping integral $t$, and the magnetic flux 
through the hexagonal cell in flux quanta $h/e$.)}
\end{figure}

Before discussing the energy spectrum of the confined  system in 
crossed  magnetic and electric fields, it is useful  to remind 
some spectral peculiarities of the infinite graphene sheet in perpendicular  
magnetic field. In this case, the spectrum  is composed of two Hofstadter 
butterflies containing  both conventional Landau levels (which depend linearly 
on the magnetic field $B$) and relativistic Landau levels 
(which depend on the magnetic field as $B^{1/2}$). 
Also specific to graphene is a flat Landau level n=0 that appears in the 
middle of the spectrum (at $E=0$). 
On the other hand, in the case of the {\it finite size} graphene, 
the confinement induces edge states, which fill the interlevel gaps, 
and a slight lifting of the level degeneracy. These aspects  can be noticed 
in Fig.5a, where the eigenvalues, obtained by the numerical diagonalization 
of the Hamiltonian (1) with vanishing boundary conditions, are shown as 
function of the magnetic flux for the relativistic (low energy) range of the 
spectrum. The additional effect of the in-plane electric field can be seen 
by inspection of Fig.5b, but, an easier examination can be done by observing  
the diamagnetic moments $M_n$ of the individual states $n$ in Fig.6.  
Since the sign of $M_n$ reveals the chirality of the state $|n>$, 
the diamagnetic moment is a convenient tool for probing the bulk  and,  
respectively, the edge states, which are distinguished by their opposite 
chirality. The calculation of  $M_n$ can be easily performed following the 
recipe described in \cite{Aldea1}.

%Fig6
\begin{figure}[!ht]
\centering{
\includegraphics[scale=0.8]{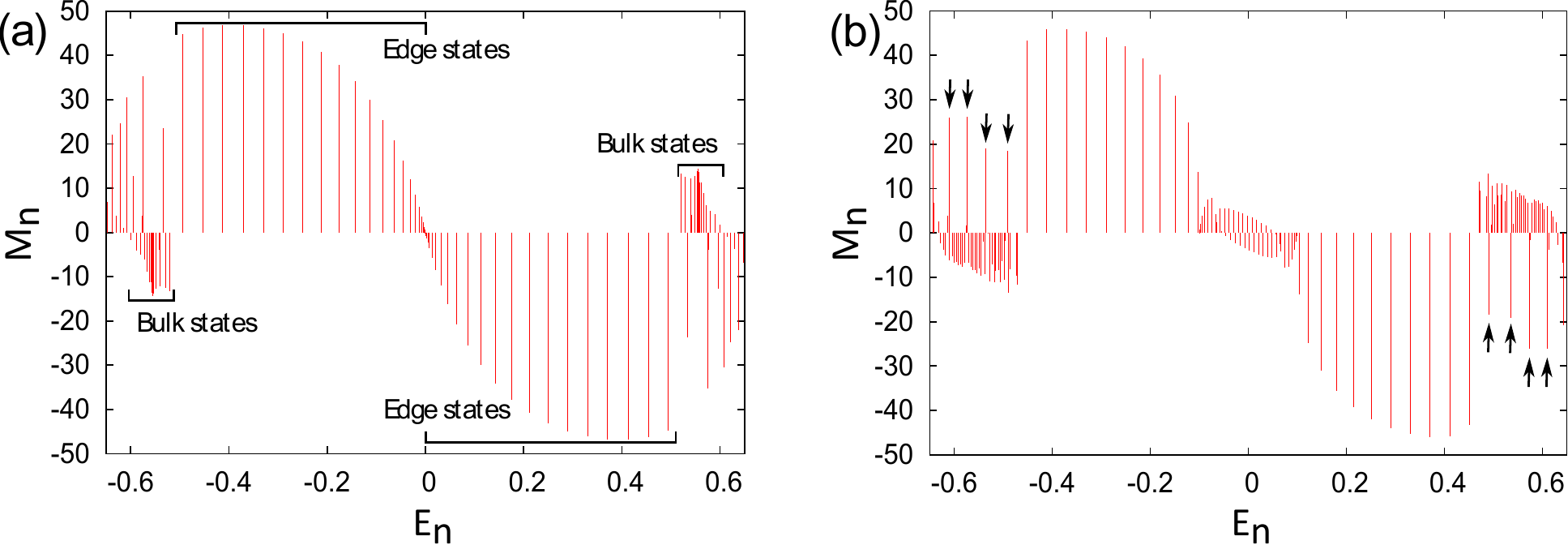}}
\centering
\caption{The orbital magnetization $M_n$ of the eigenstates versus the
eigenenergies $E_n$ for a graphene plaquette of dimension $41 \times 40$, with 
no electric field (a), and with an in-plane electric field $e\mathcal{E}L_y/t=0.2$ (b),
$\phi/\phi_0=0.03, \beta=0.03$. 
In the panel (b), the {\it shortcut edge states} located in the first relativistic 
band (see text) are marked with arrows. Alternate magnetizations are to be noticed 
in the zero-energy range. }
\end{figure}

The following aspects can be noticed by comparing the two panels in Fig.6:\\
i) the  energy range occupied by the bulk states at  $\mathcal{E} \ne 0$ is more
extended than in  the case  of vanishing electric field.
This is understandable since the electric field lifts the quasi-degeneracy of any
Landau band, giving rise to a Stark fan with increasing field.\\
ii) the presence among the bulk states at $\mathcal{E} \ne 0$ of several
states of reverse chirality, mentioned with arrows in Fig.6b~;
their chirality indicates that the states are of edge-type, nevertheless 
they  behave unusually.  Indeed, the calculation of the charge density 
$|\Psi(\vec{r})|^2$ proves that all  these states keep the localization 
along the edges, but exhibits also a {\it ridge} in the middle of the plaquette,
which is  perpendicular on the direction of the electric field, and shortcuts two 
opposite sides.
For a given magnetic field, the position of the ridge on the plaquette depends on the
energy $E_n$ and  can be modified by changing the electric field.
Such states generated by the application of the electric field,
are shown in Fig.7, and  will be called {\it shortcut edge states}. 

In the classical picture, the normal edge states are assimilated to
skipping cyclotron orbits along an equipotential line near the hard walls.
In the same line of thinking,  one may assume that the applied electric bias creates 
an internal
barrier that limits the electron motion and compels the skipping orbit to close
along the equipotential line along the electric barrier. This would be the
classical picture of a shortcut edge state.

The diamagnetic current carried by a shortcut edge state flows along the 
plaquette edges, but part of it closes through the middle. When leads are attached 
to the plaquette, the shortcut obviously affects the electron transmittance between 
the different leads, with consequences for the quantum Hall effect that will be 
discussed in sec.III.

One cannot  disregard the presence also in  the left panel of Fig.6
(at $\mathcal{E}=0$) of a state of  reversed  chirality in the first Landau band. 
We checked it by calculating the charge distribution, and it turned  out 
that the state is a usual edge state going around all the four edges of the sample. 
We assume that it appears  accidentally among 
the bulk states as a finite size effect for the given plaquette,
at the given magnetic field, due to some specific hidden symmetry.

At this stage one  has to discuss the question of the electric field 
strength. It is known from the continuous approximation used in \cite{Baskaran, Castro} 
that, with increasing value of the parameter $\beta$, the gaps
%$\beta=\mathcal{E}/v_F B$ ($v_F$=Fermi velocity), the gaps
between the relativistic Landau levels diminish, and there is a critical value 
$\beta=1$ at which all the gaps vanish.  We need to say that our results, 
obtained in the finite lattice model are generated for  values of $\beta$
which are lower than those used in Lukose {\it et al} \cite{Baskaran}
and much lower than the critical $\beta$. 

The  closing of the gaps  with increasing  electric field obviously occurs also 
in the lattice model due to the broadening of the Dirac-Landau bands provided by 
the above mentioned formation of the Stark fan \cite{Note-critic}. 
In this way, the gaps disappear together with the  normal edge states located inside.
However, an interesting question concerns the fate of the shortcut edge states at
high values of $\beta$, beyond the gap closure. One has to remember (see Fig.6b) that
the shortcut edge states appear in the bands (not in the gaps), so that one may suppose
that they will survive even at high values  of $\beta$, when the gaps disappear and the
mixing of the adjacent bands occurs. This situation, if true, should be seen
in the quantum Hall effect,  which is no more carried by the normal edge states, but
only by the shortcut edge states. Indeed, the calculation of the Hall resistance 
shows in Fig.11 that for $e\mathcal{E}L_y/t=0.5$
the normal plateaus disappear (except the first one corresponding to the largest 
first gap, which is not yet closed), but the intermediate plateaus at $R_H=2/3$ 
and $R_H=4/16$, supported by  the shortcut edge states,  still exist.

%Fig7
\begin{figure}[!ht]
\centering{
\includegraphics[scale=0.8]{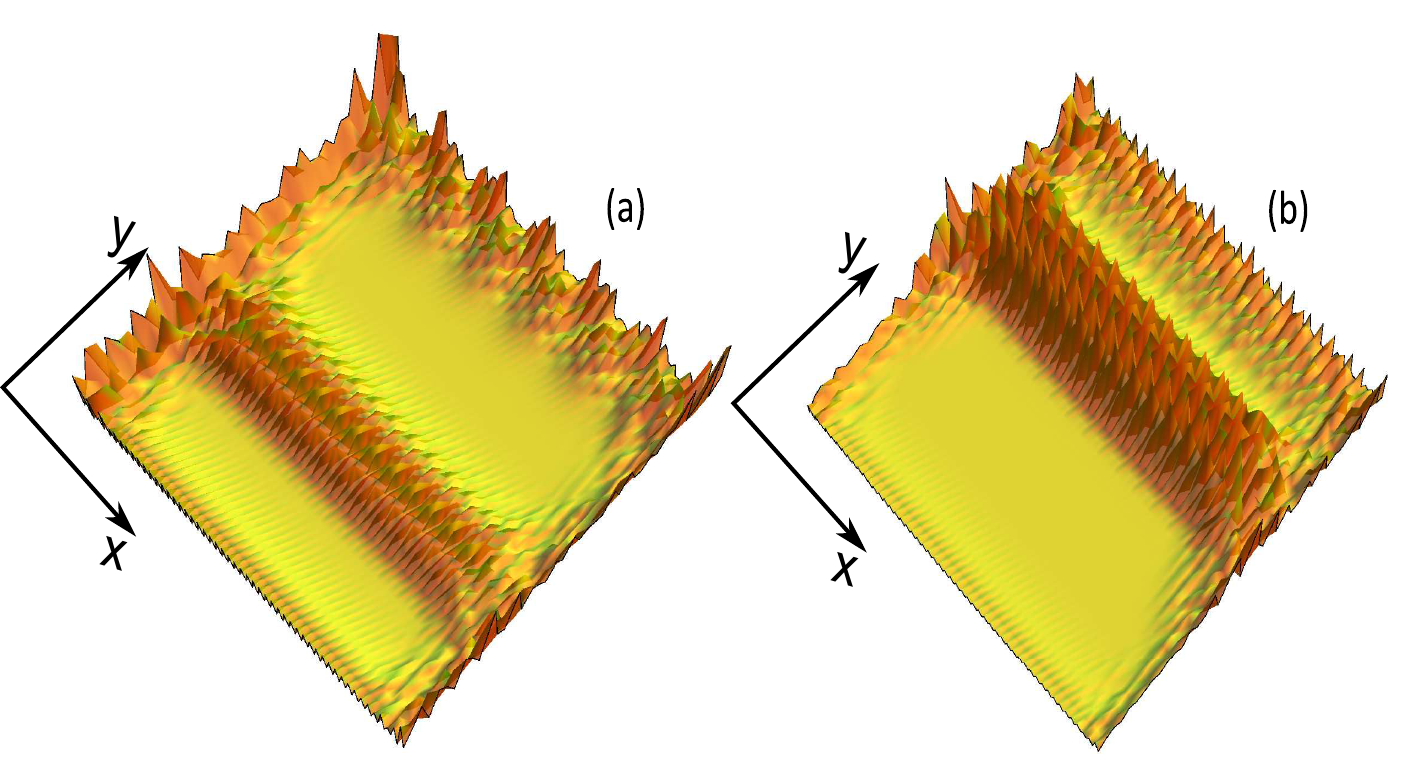}}
\centering
\caption{$|\Psi(\vec{r})|^2 $ for two shortcut edge states from
the first relativistic Landau band corresponding to the energies $E=0.5306$ (panel a)
and $E=0.5786$ (panel b).
The number of lattice sites is $105\times40$, 
$\phi/\phi_0=0.03$, and $e\mathcal{E}L_y/t=0.2$, corresponding to $\beta=0.03$.}
\end{figure}

One may also ask the question whether 
the shortcut edge states are present at any small electric field
or there is a threshold above which these states appear. 
The question is conceptually interesting
since it can give a physical hint, beyond the numerical result, for the formation of the
shortcut edge states. 
The answer is based on the discrete character of the energy spectrum of the
finite plaquette and on the different response of the two types of states (band and edge)
to the presence of the electric field. As we already noticed, the band states are linearly
shifted on the energy scale (giving rise to the Stark fan), while the edge-type energies are
robust. Consequently, there is an $\mathcal{E}$ for which a bulk state becomes 
resonant with the
first edge state met in the gap. Up to this value of the electric field the separate character 
of the edge and bulk states is maintained \cite{Note-shape}, however they hybridize 
with each other at the resonance.
It turns out that the new state is a shortcut edge state, as it is
proved by the numerical calculation of the charge density distribution on the 
plaquette.
So, one concludes that for the existence of the shortcut edge states 
the electric field should be higher
than a minimal value determined  by the edge states level spacing (call it $\delta$), 
i.e. $e\mathcal{E} L > \delta$,
where $L$ is the length of the sample in the direction of the field.
Obviously, $\delta$ depends on the plaquette dimension and is
different in different gaps. 
Since $\delta$ vanishes in the limit $L\rightarrow \infty$, 
the critical electric field
goes to zero in the limit of large systems.
In the case of the plaquette considered in Fig.6
(41$\times$40 sites corresponding to 4.9$\times$8.4 $nm^2$  and  $\phi/\phi_0=0.03$), 
the estimated critical 
$\beta$ is $0.006$ in the first gap. 

\subsection{The  special case around the energy $E=0$}
One may notice in  Fig.5b that the additional in-plane electric field lifts 
the degeneracy of the n=0 Landau level and gives rise to a broad splitting 
in the energy range (-0.1, 0.1). Obviously, the splitting depends on the strength
of the applied electric field.
The resulting states that fill this range appear in pairs of opposite 
chirality \cite{Note1}, fact that can be observed either looking at the 
derivative $dE_{n}/d\phi$ in Fig.5b or at the sign of the magnetic moments 
in Fig.6b. 
The small values of the magnetic moments in the energy range close to $E=0$ 
suggest that the surface encircled by the diamagnetic currents 
might be also small. In order to check this idea, we calculated the charge 
distribution $|\Psi_{n}(\vec{r})|^2 $ on the plaquette, 
and a somewhat surprising result came out. 
Namely, all the states  shortcut the plaquette, looking as in Fig.8, 
where two  eigenstates, consecutive  on the energy scale, are shown.
One observes that they  encircle complementary areas on the plaquette, 
which are controlled by the electric field intensity.
This behavior, regarding both the chirality and the arrangement on the plaquette,
is assigned to  all states that appear in the central energy range.  

The states described above represent a second kind of shortcut edge states to be
found in the finite size graphene plaquette. 
We expect them to give new features to the transport properties near the zero-energy.
We advance already the idea that the asymmetric positioning of the states on the
plaquette should manifest itself in the dependence of the transport properties 
on the configuration of the current and voltage leads in the Hall device. 
The topic will be explored in the next section.

%Fig8
\begin{figure}[!ht]
\centering{
\includegraphics[scale=0.9]{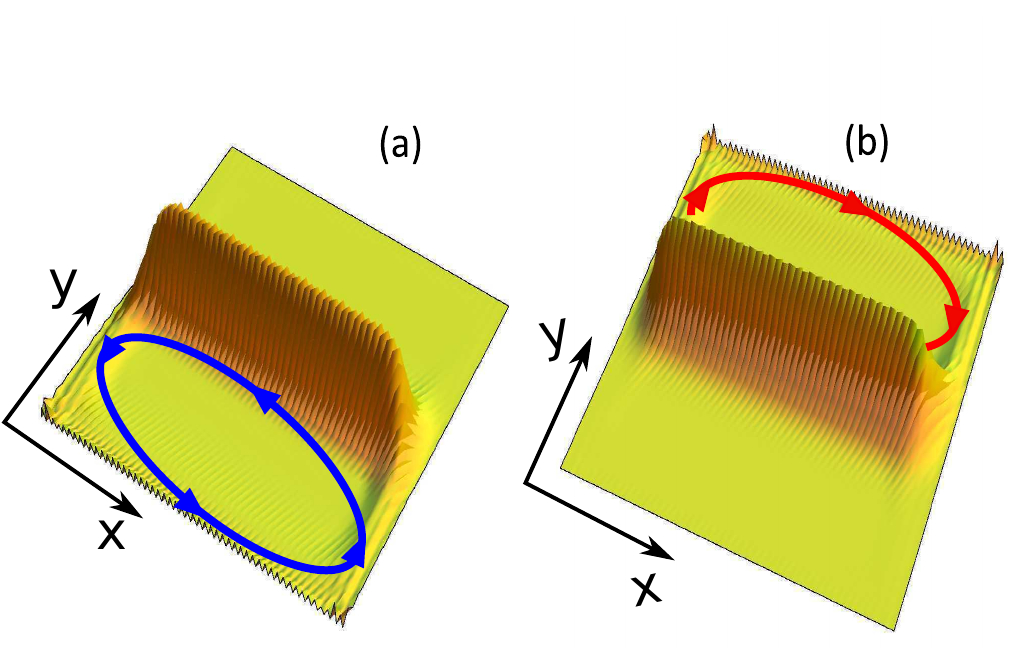}}
\centering
\caption{(Color online) $|\Psi(\vec{r})|^2 $ for two consecutive shortcut 
edge states from the central Landau band (n=0) in the presence of the
in-plane electric field. The states show  alternate chiralities (magnetizations) 
which are illustrated by blue and red loops. The number of lattice sites is 
$105\times40$, $\phi/\phi_0=0.03$, $e\mathcal{E}L_y/t=0.2$ and $\beta=0.03$.}
\end{figure}

\section{Peculiar  IQHE of the graphene plaquette in the presence of the 
in-plane electric field}
In order to reveal the specific transport proprieties of the graphene plaquette 
in crossed electric and magnetic fields, we simulate a quantum Hall device  
by attaching four leads to  the  graphene plaquette. 
In this way, we investigate  the modifications produced by the in-plane electric 
field to the already familiar picture of the IQHE in graphene, which in the 
relativistic energy range shows plateaus at $R_H=\pm \frac{h}{e^2} \frac{1}{2n+1}$  
(with $n=0,1,2,..$, and forgetting about the spin degeneracy).
The basic idea is that the novel  edge states, due to their specific properties 
induced by the applied bias (and discussed in the previous section), affects 
the electron transmittance between different leads, and, implicitly, the plateaus 
of the quantum Hall resistance.

It is of interest to note that the   quantum Hall behavior  in the 
two distinct domains occupied by the {\it shortcut edge states} can 
be guessed heuristically by exploiting the qualitative features of the edge states 
and the way in which they shortcut the plaquette and interconnect the leads. 
This approach suggests the presence of some new, specific quantum Hall plateaus,
 which, however, should be  checked numerically.

The numerical calculation is performed in the Landauer-B\"{u}ttiker formalism,
the basic formulas being reminded here.
In a four lead-device the charge current through the lead $\alpha$
can be written in the linear regime as: 
\begin{equation}
I_{\alpha}=\sum_{\beta=1}^4~g_{\alpha\beta} V_{\beta},  
\end{equation}
where 
$g_{\alpha\beta}$ is the conductance matrix,
$V_{\beta}$ is the potential at the contact $\beta$, and the lead indices 
$\alpha,\beta= 1,...,4$. 
The current conservation and the possibility to choose 
arbitrarily the origin of the potential impose the conservation  rules:
\begin{equation}
\sum_{\alpha=1}^4 g_{\alpha\beta}=\sum_{\beta=1}^4 g_{\alpha\beta}=0~ .
\end{equation}

For  $\alpha\ne\beta$, the conductance  $g_{\alpha\beta}$ can be expressed
in terms of the transmission coefficients $T_{\alpha\beta}$ between the leads
$\alpha$ and $\beta$ as 
%In terms of transmission probabilities $T_{\alpha\beta},
%for $\alpha\ne\beta$,
 $g_{\alpha\beta}= \frac{e^2}{h} T_{\alpha\beta}$ .
%while the diagonal term $g_{\alpha\alpha}$ can be calculated either
On the other hand, the diagonal term $g_{\alpha\alpha}$ can be obtained either
from the above conservation law or using the recipe 
$g_{\alpha\alpha}= {\frac{e^2}{h}} (T_{\alpha\alpha}-M^c_{\alpha})$,
where $M^{c}_{\alpha}$ is the number of channels in the lead $\alpha$. 
%The second  recipe is an immediate outcome of the Datta's formalism in 
The second  recipe can be  immediately  deduced from the Datta's formalism in 
\cite{Datta}.

The transmission coefficients $T_{\alpha\beta}$   
can be calculated using the  Green function approach.  
The method pretends to know the full Hamiltonian consisting of the sample 
Hamiltonian $H^S$ given in our case by Eq.(1) and supplementary terms, 
which describe the leads $H^L$ and the sample-lead coupling  $H^{SL}$:
\begin{equation}
H=H^{S}+H^{L}+\tau H^{SL}.
\end{equation} 
We consider many-channel perfect leads similar to those introduced 
in \cite{Buettiker},
the strength of the lead-sample coupling used in the numerical calculation 
being $\tau=2 t$. 
The role of the leads is to inject and collect the current flowing 
through the graphene plaquette;  
in the considered tight-binding model, each lead consists of  $M^c$
semi-infinite one-dimensional conducting chains, 
which are attached to consecutive sites of the plaquette. 
In terms of creation (annihilation) operators acting on the lead sites,
the Hamiltonian $H^L$ reads \cite{Aldea}:
\begin{equation}
H^L=\sum_{\alpha}H^L_{\alpha},~~~
H^L_{\alpha}=t\sum_{\nu=1}^{M^c} \sum_{n\ge 1}c^{\dagger}_{\alpha,\nu,n}
c_{\alpha,\nu,n+1} +H.c.~ ,
\end{equation} 
where $\nu$ counts the chains, 
$n$ counts the sites along  any
semi-infinite chain, and $t$ is the hopping integral on the chain.

The transmission coefficient $T_{\alpha\beta}$, which describes 
the electron propagation {\it from the lead $\beta$ to the lead $\alpha$}  
at the Fermi energy $E_F$, is given by the  expression: 
\begin{equation}
T_{\alpha\beta}(E_{F})=4\tau^{4} \sum_{\nu,\nu'}
|G_{\alpha\nu,\beta\nu'}^{+}(E_{F})|^{2}Im g_{\alpha,\nu}^{L}(E_{F})
Im g_{\beta,\nu'}^{L}(E_{F}),~~\alpha\ne\beta,
\end{equation} 
where $G^+$ is the retarded Green function of the system in the presence of
the coupled leads, and $g^L$ is the lead Green function  (so that 
$Im g_{\alpha,\nu}^{L}(E_{F})$ represents the density of states at the Fermi energy
of the chain $\nu$ in the lead $\alpha$). 

Next, if $V_{\alpha\beta}(\alpha\ne\beta)$ is the voltage drop measured between 
the contacts $\alpha$ and $\beta$ when the current $I_{\delta \gamma}$ 
flows between the contacts ($\delta, \gamma$), with the notation 
$R_{\gamma \delta, \alpha \beta}=V_{\alpha \beta}/I_{\delta \gamma}$
(introduced by van der Paw \cite{Paw} for many terminal devices),
the transverse (Hall) resistance of the four-terminal device sketched in Fig.9
is given by:
\begin{equation}
R_H=(R_{13,24}-R_{24,13})/2= 
(g_{23}g_{41}-g_{21}g_{43}-g_{32}g_{14}+g_{12} g_{34})/2D, 
\end{equation} 
where $D$ is a $3\times 3$ subdeterminant of the conductance matrix defined 
in Eq.(6).  We note that, since the conductance matrix satisfy the
conservation rules Eq.(7), all the $3\times 3$ subdeterminants  are equal 
(up to a sign which is irrelevant here) and different from zero.

%Fig9
\begin{figure}[!ht]
\includegraphics[scale=1.0]{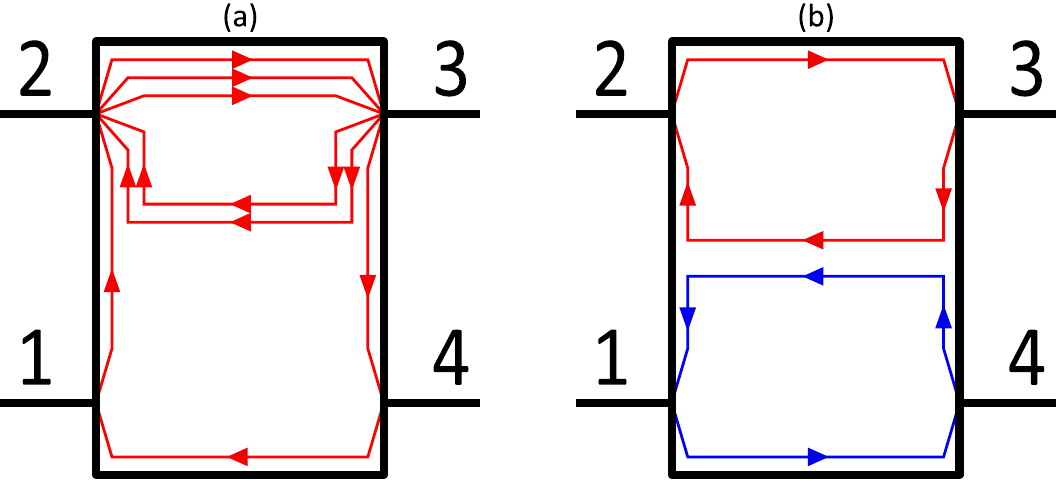}
\caption{(Color online) The sketch of the current carrying channels 
when four leads are attached to the graphene plaquette.
The involved shortcut edge states are those from Fig.7 in the panel (a)
and those from Fig.8 in the panel (b).}
\end{figure}

The  {\it first kind} of shortcut edge states was identified in Fig.6b 
as being dispersed among the bulk states in  the  relativistic Landau band 
broadened by the electric field. Then, we may argue that, 
instead of observing the expected
drop of the Hall resistance between two consecutive plateaus, one may find 
an uncommon plateau supported by the new type of edge states located here. 
Such a  state looks like in Fig.7, while  the  way it  interconnects the leads 
is shown in Fig.9a.  The distribution of the channels and their chirality 
(shown in Fig.9a for a given direction of the magnetic field) help in 
specifying $T_{\alpha\beta}$. For instance, one notices that 
the leads 1 and 2 are 
interconnected by one channel, while the leads 2 and 3 are interconnected by  
three channels, meaning that $T_{21}=1, T_{32}=3, T_{12}=0$, etc. 
Then, using also the  conservation rules mentioned above,
the whole 4$\times$4  conductance matrix {\bf g}  
can be easily built up as:
\begin{equation}
\text{\textsl{\bf g}}=\frac{e^2}{h}
\begin{pmatrix}
  -1 & 0 & 0 & 1 \\
  ~ 1 &-3 & 2 & 0 \\
  ~ 0 &~ 3 & -3 & 0  \\
  ~ 0 &~ 0 & 1 & -1 
\end{pmatrix}.
\end{equation}

The Hall resistance $R_H$ can be calculated now
according to the Landauer-B\"{u}ttiker recipe  Eq.(11), the result being 
$R_H=2/3 (h/e^2)$. This value represents a new plateau  placed 
in the relativistic domain at half the distance between the usual plateaus 
$R_H=1(h/e^2)$ and $R_H=1/3(h/e^2)$. This plateau is clearly evidenced by
the numerical calculation  in Fig.11.
(We remind once again that the spin degeneracy is not taken into account.) 

The {\it second kind} of shortcut edge states, resulting from the 
degeneracy lifting of the n=0 Landau level, are located in the middle 
of the spectrum. When calculating numerically the quantum Hall resistance, 
we noticed in this range some unusual features of the conductance matrix, 
which are listed below:
\begin{subequations}
\label{allequations} % notice location
\begin{eqnarray}
T_{12}&=&T_{21}=0,~~T_{34}=T_{43}=0,\label{equationa}
\\
T_{41}&=&T_{32}=1, \label{equationb}
\\
T_{41}&=&T_{14}+T_{13},~~ T_{32}=T_{23}+T_{24}.\label{equationc}
\end{eqnarray}
\end{subequations}

These relations are quite different from the usual conditions for the 
realization of the conventional IQHE, which (for a single channel and  
a given orientation of the perpendicular magnetic field) read simply
$T_{\alpha,\alpha+1}=1$ and $T_{\alpha+1,\alpha}=0$   
(for any lead $\alpha$), all the other elements being zero.
The Eq.(13a) and Eq.(13b) are corroborated by the properties of the 
edge states already observed in Fig.8, where the two states cover 
complementary areas of the plaquette and have opposite chirality. 
These  features of the quantum states suggest the configuration of 
the channels on the plaquette  sketched in Fig.9b.
One may notice that the contacts 1 and 2 are not connected, 
meaning that both $T_{12}$ and $T_{21}$ vanish (similarly,  
$T_{34}=T_{43}=0$), and that the blue  state connects the leads 1 
and 4 in the direction  providing $T_{41}=1$  
(similarly, $T_{32}=1$ for the red state).

However,  we are still left with the intriguing relation 
$T_{41}=T_{14}+T_{13}$, where $T_{13}\ne 0$,  although the contacts 1 and 3 
are apparently disconnected as in  Fig.9b.
Thus, the transmission coefficient  $T_{13}$ breaks the expected symmetry 
$T_{14}=T_{41}$, and plays the role of  a 'leakage'  between the two 
(red and blue) circuits.
A plausible explanation for this effect might be  that the  states, 
being very close on the  energy scale, can  be hybridized easily
by the perturbation introduced in the system by the lead-plaquette 
coupling $\tau$ (see Eq.(8)). 
In this way, the two circuits become interconnected if 
$\tau \sim \Delta $, where $\Delta$ is the mean interlevel distance 
in the corresponding energy range. 
Then,  using this conjecture, with the notation 
$T_{13}=T_{24}=\delta$, and  observing in  Fig.9b the manner  
the  contacts are bridged, the corresponding matrix reads:
\begin{equation}
\text{\textsl{\bf g}} =\frac{e^2}{h}
\begin{pmatrix}
  -1 & ~0 & \delta & 1-\delta \\
   ~0 & -1 & 1-\delta &~ \delta \\
   ~0 & ~1 & -1 &~ 0  \\
   ~1 & ~0 & ~0 & -1 
\end{pmatrix}.
\end{equation}

%Fig10
\begin{figure}[!ht]
\includegraphics[scale=0.7]{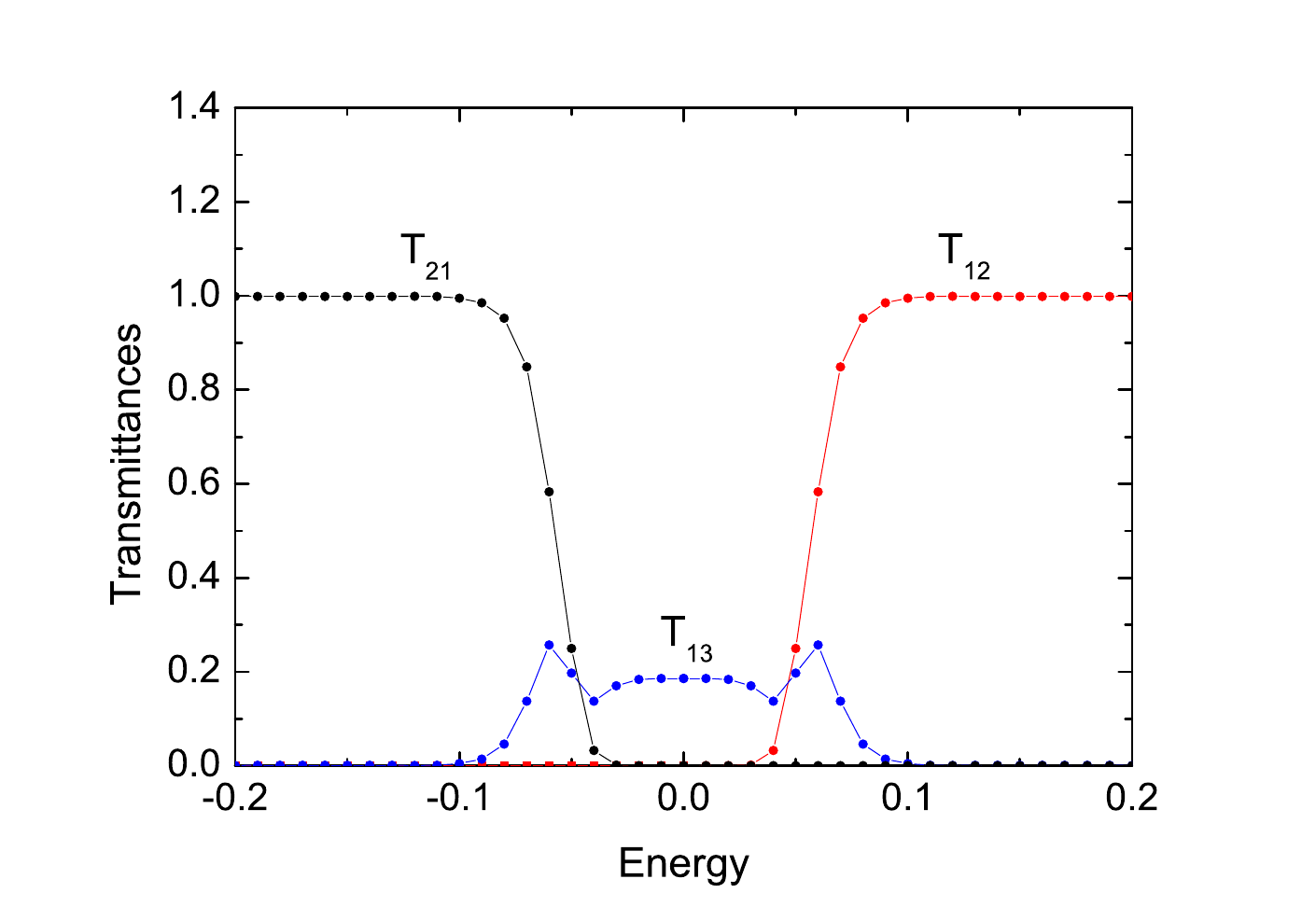}
\caption{(Color online) Transmission coefficients vs. energy for a 
mesoscopic graphene in the presence of both perpendicular magnetic and 
in-plane electric fields. 
Around the zero-energy, $T_{21}=T_{12}=0$  and $T_{13}\ne 0$ evidencing 
the 'leakage' between the two circuits illustrated in Fig.9b. The number 
of lattice sites is $51\times 100$, the magnetic flux $\phi/\phi_0=0.03$, 
electric field $e\mathcal{E}L_y/t=0.25$ and $\beta=0.015$.}
\end{figure}

By using again Eq.(11), we get this time the plateau $R_H=0$ that should 
become visible in the middle of the spectrum about the energy $E=0$. 
This value obtained by the  heuristic method is also confirmed by the  
numerical calculation in Fig.11. 
The numerical values obtained 
%according to Eq.(3) 
for $T_{12}$, $T_{21}$ 
and $T_{13}$ are shown in Fig.10. The unusual behavior can be noticed indeed
about $E=0$, where both $T_{12}$ and $T_{21}$ vanish, while  $T_{13}$  
is different from zero. Outside this energy range, $T_{13}$  vanishes  
(and also $T_{42}$), while $T_{12}$ and $T_{21}$ take the normal values 
that characterize the first relativistic gaps in  the graphene Hofstadter 
spectrum.

The  result of the numerical investigation of the Hall resistance  as function 
of the energy in the relativistic range of the spectrum, in both the presence 
and  absence of the applied electric bias, is shown in Fig.11. 
At vanishing bias, the red curve exhibits the well-known plateaus of the IQHE 
in graphene at $1, 1/3, 1/5$ (in units $h/e^2$). 
However, at $\mathcal{E}\ne 0$, the blue curve shows supplementary unconventional 
plateaus that appear in-between at $2/3$ and $4/15$, and also a very specific 
plateau  at $R_H=0$.
%\vskip-2cm
%Fig11
\begin{figure}[!ht]
\includegraphics[scale=1.2]{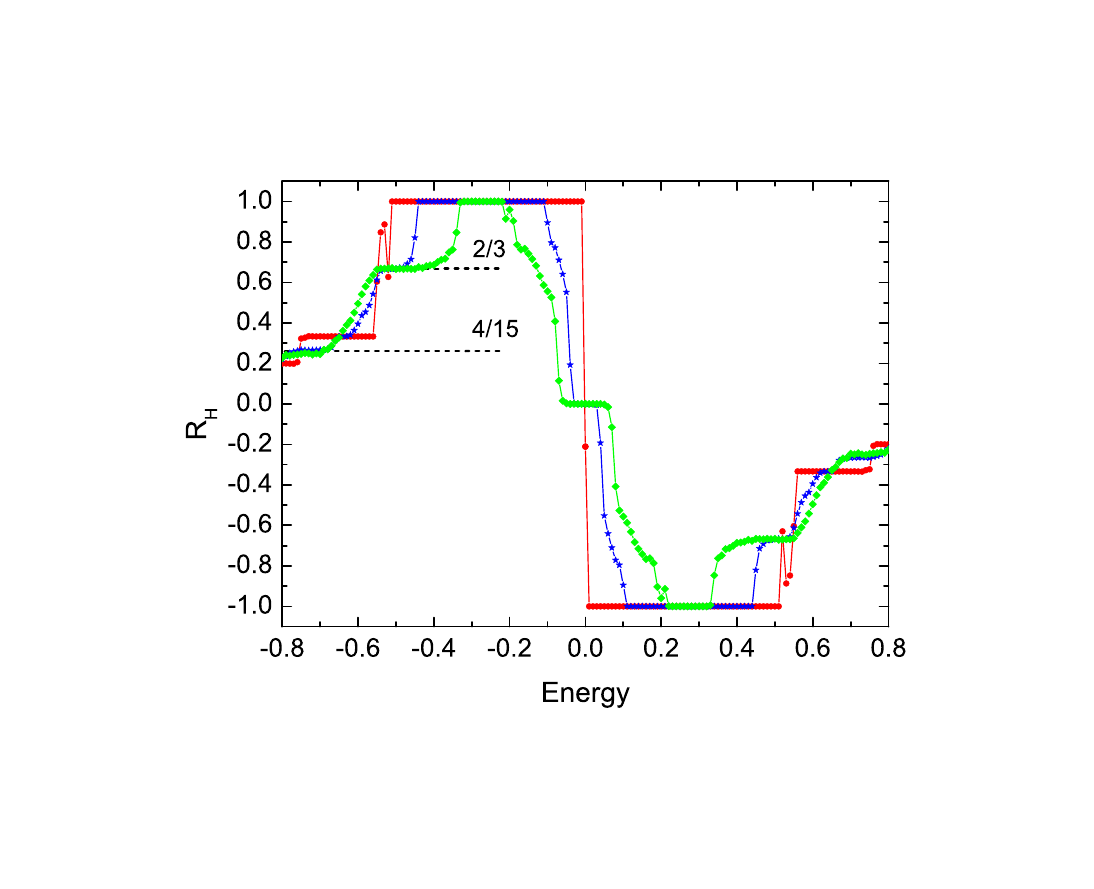}
\vskip-2.5cm
\caption{(Color online) Hall resistance as a function of energy with no electric 
field (red curve) and with electric filed $e\mathcal{E}L_y/t=0.25$ (blue) 
and 0.5 (green). We notice the appearance of the intermediate Hall plateaus 
in the presence of the electric field.
At strong $\mathcal{E}$ (green curve) the normal plateaus at 
$R_H=1/5$ and $1/3$ disappear due to closing of the corresponding gaps in 
the spectrum.}
(The number of lattice sites $51\times100$ and $\phi/\phi_0=0.03$, $\beta=0.015$
(blue) and $\beta=0.03$ (green).)
\end{figure}

In what concerns the longitudinal resistance, an interesting result comes from 
the possibility to define (and measure) it in two different ways, 
depending on the chosen configuration for the voltage and current terminals, 
namely:
\begin{eqnarray}
R_L^{A}=(R_{23,14}+R_{14,23})/2=
(g_{12}g_{43}-g_{13}g_{42}+g_{21}g_{34}-g_{24}g_{31})/2D , \\
R_L^{B}=(R_{12,34}+R_{34,12})/2= 
(g_{31}g_{42}-g_{32}g_{41}+g_{13}g_{24}-g_{14}g_{23})/2D . 
\end{eqnarray}

Using again  Eq.(14), valid in the energy range near $E=0$, 
one finds  that the longitudinal resistance shows a dissipativeless behavior  
$R_L^A=0$ in  the configuration $A$, but gets a finite value, which depends on 
the 'leakage' parameter $\delta$, in the configuration $B$, 
namely $R_L^B=(1-\delta)/\delta$. 

Since the numerical calculation refers to the Hall resistance of the 
{\it finite} graphene plaquette, it is opportune here to make  some 
comments concerning the effect of the finite size on the structure 
of the Landau bands. 
The discussion is prompted by the fact that the Hall resistance
described by the red curve in Fig.11 does not drop abruptly (as one may expect)
between the plateaus $R_H=1$ and $R_H= 1/3$, separated by the first
Dirac-Landau band, but even shows  an oscillation. 
The confinement represents a perturbation which, 
on one hand, introduces edge states in the spectrum, and, on the other hand,
slightly lifts the degeneracy of the Landau states 
giving rise to a small band  broadening.
Obviously, as in any mesoscopic problem, the effect of the surface depends 
on the dimension of the plaquette, and , more precisely, the smaller 
the plaquette, the larger the broadening of the Landau bands is.
However, the splitting occurs differently at different magnetic fluxes,
so that, the overall aspect of the broadened band is a bunch of states 
exhibiting numerous anticrossing points when the magnetic field is varied.
Our preliminary observations, based on the numerical calculation of the 
Hofstadter spectrum, indicate that the relativistic Landau bands, 
placed in the central part of the spectrum, are more sensitive to the 
confinement then the conventional Landau bands, placed at the spectrum 
extremities. This discussion  is actually  beyond the aim of the present 
paper and we only want to give a hint for possible unanticipated behavior 
of $R_H$  between the quantum plateaus.

As a last comment, we observe that in the presence of the electric bias, 
over the slight broadening due to the confinement, 
the broadening produced by the electric field is added 
(the so-called Stark ladder). Then, the transition between the successive
plateaus of the Hall resistance becomes more gradual, as it can be noticed 
in Fig.11 (blue and green curves).

\section{Summary and conclusions} 
In conclusion, we have proved the presence of {\it shortcut edge states}
in the relativistic energy range of the mesoscopic graphene 
subjected to crossed  magnetic and electric fields, 
and their influence on the transport properties denoted by new, unconventional,
plateaus of the quantum Hall resistance.
The novel states are defined as  being only partially extended along the  edges, 
and getting closed through the middle of the plaquette, the shortcut  being 
controlled by the electric field.

In order to get  insight on the emergence of the electrically induced shortcut 
states, it was helpful to study first the case of  the zig-zag graphene ribbon.
By applying the electric field   perpendicularly to the zig-zag edges, we found 
that the Landau levels get tilted and, as a consequence, some of the edge channels
located along the zig-zag edges are pushed by the 
electric field into the middle of the ribbon (as in  Fig.3b). 
Interestingly, even the zero energy  band states respond  to the electric field, 
turning into current carrying states, some being located near the edges, 
and  others in  the middle of the  ribbon (as in Fig.3c). 
For the graphene ribbon, we prove the scaled dependence of the
low-energy Dirac-Landau spectrum on the external parameters 
$BL_y$ and $\mathcal{E} L_y^{3/2}$ in Eqs.(4-5).

In the case of the mesoscopic graphene plaquette, we found two kinds of 
shortcut edge states located differently in the relativistic range of 
the Hofstadter spectrum. Some of them are dispersed among the bulk states 
in the relativistic Landau bands, while the second type of such states 
arises in the middle of the spectrum due to the splitting of the n=0 Landau 
band induced by the electric field.
In order to determine the chirality of the states, we calculated the 
diamagnetic moment of all quantum states in the relativistic range,
with and without electric field, as shown in Fig.6.

The shortcut of the edge states created by the  electric field modifies the 
conductance matrix in the four-lead Hall device, and gives rise to novel
plateaus of the  quantum Hall resistance. We presented heuristic arguments 
and numerical calculations for identifying the position of the new plateaus.
The shortcut edge states of the first kind generate intermediate  plateaus 
between the usual ones at 
$R_H=\pm \frac{1}{2}(\frac{1}{2n+1}+\frac{1}{2n+3})$ with n=0,1,2... (in
$h/e^2$  units for the spinless case).

In the central part of the spectrum, the second kind of shortcut edge states 
come into play. They appear in pairs with different chiralities and exhibit 
current loops that  encircle complementary areas of the plaquette, 
as being depicted in Fig.8. These specific properties generate
the plateau $R_H = 0$, and  dissipative or non-dissipative behavior of the 
longitudinal resistance $R_L$, depending on the leads configuration.

The presented results put forward a mechanism for manipulating the transport 
channels in the quantum Hall regime  by using an in-plane electric bias, 
and extend the understanding  of the edge states in the relativistic energy 
range of the mesoscopic graphene.

\section{Acknowledgments}
We acknowledge  support from PNII-ID-PCE Research Programme
(grant no 0091/2011) and Core Programme (contract no.45N/2009). One of the
authors (AA) is grateful to Achim Rosch for the support in the frame
of Kernprofilberech QM2 at the Institute for Theoretical Physics, 
University of Cologne.

\appendix*
\section{}
As we already observed  in Sec.II A, the eigenenergies 
of the graphene ribbon in crossed electric and magnetic fields
have the symmetry
$E_{k^0_x+k_x}=-E_{k^0_x-k_x}$ around a reflecting point $k^0_x$. 
In order to prove this property and find the value of $k^0_x$, it is 
appropriate to express the Hamiltonian in terms of the Fourier transforms 
of the creation and annihilation operators  as in \cite{Neto}:   
\begin{eqnarray}
H=\sum_{\substack{k_x,m}}{\epsilon^{a}_m a^{\dag}_{k_x,m}a_{k_x,m}+
\epsilon^{b}_m b^{\dag}_{k_x,m}b_{k_x,m}+
\textit{t}(e^{i\phi(m)}a^{\dag}_{k_x,m}b_{k_x,m}} 
\nonumber \\ 
+e^{i\phi(m)}e^{-ik_{x}}b^{\dag}_{k_x,m}a_{k_x,m}+b^{\dag}_{k_x,m+1}a_{k_x,m}+H.c.), 
\end{eqnarray}
where $m$ labels the unit cells along the Oy-direction (m=1,..,M). 
The magnetic field manifests itself in the Peierls phases $\phi(m)$, 
while the electric field enters the atomic energies $\epsilon^{a,b}_m$. 
Applying the electric bias symmetrically on the width of the ribbon, 
and  using the known inversion symmetry with respect to the middle of the 
hexagonal lattice (which moves the
atoms A in the atoms B and vice-versa), the atomic energies satisfy the relation:
\begin{equation}
\epsilon^{a}_{M+1-m}=-\epsilon^{b}_m.
\end{equation}
Let us measure the momentum $k_x$ from a not yet specified  origin $k^0_x$,
and let  the function:
\begin{equation}
\Psi_{k_x^0-k_x}= \sum_m \alpha(k^0_x-k_x,m) a^{\dagger}_{k^0_x-k_x,m}|0>+
 \beta(k^0_x-k_x,m) b^{\dag}_{k^0_x-k_x,m}|0>
\end{equation}
be an eigenfunction of the Hamiltonian, such that:
\begin{equation}
H \Psi_{k^0_x-k_x}= E_{k^0_x-k_x}\Psi_{k^0_x-k_x}.
\end{equation}
The objective is now to find an unitary operator $P$ with two properties:\\
i) to anticommute with $H$, i.e., ~$PHP^{-1}= -H$,\\
ii) to move the function $\Psi_{k^0_x-k_x}$ into another function depending 
on $k^0_x+k_x$, i.e.,
~$P\Psi_{k_x^0-k_x}=\tilde\Psi_{k_x^0+k_x}$.\\
Then, applying the operator $P$ to the left of Eq.(A.4), and using the first 
property, one gets obviously:
\begin{equation}
H\tilde\Psi_{k_x^0+k_x}= -E_{k^0_x-k_x} \tilde\Psi_{k_x^0+k_x}, 
\end{equation}
meaning that $E_{k^0_x+k_x}= -E_{k^0_x-k_x}$.

Taking advantage of the  mentioned inversion symmetry of the lattice, 
we may guess the operator $P$ as:
\begin{eqnarray}
P=\sum_{\substack{k_x,m}}{e^{-i\lambda^{b}(k^0_x-k_x,m)} b^{\dag}_{k^0_x-k_x,M+1-m} 
a_{k^0_x+k_x,m}}
\nonumber \\
-\sum_{\substack{k_x,m}}{e^{-i\lambda^{a}(k^0_x-k_x,m)} a^{\dag}_{k^0_x-k_x,M+1-m} 
b_{k^0_x+k_x,m}}.
\end{eqnarray}
The phases $\lambda^{a,b}$ and the reflecting point $k^0_x$ result from the 
condition $[H,P]_+=0$. After lengthy but straightforward calculations one obtains
 the value $k^0_x=\phi(M)+\phi(1) $ (modulo $\pi$), which depends on the
magnetic field $B$ and the width of the ribbon $M$.

It is interesting to notice that  performing  the inverse Fourier transformation 
of the operators in  Eq.(A.6), namely $a^{\dag}_{k_x,n}=
\frac{1}{\sqrt{N}}\sum_n e^{ik_x n} a^{\dag}_{n,m}$ 
(and similarly for the $b$ operators), one obtains an operator $\mathcal{P}$ which
acts in the direct space as an inversion operator with respect to middle of the sample: 
\begin{eqnarray}
\mathcal{P} a^{\dag}_{n,m}|0>&=& e^{-i(2n+m)k^0_x} b^{\dag}_{N+1-n,M+1-m}|0> 
\nonumber \\
\mathcal{P} b^{\dag}_{n,m}|0>&=&- e^{-i(2n+m-1)k^0_x} a^{\dag}_{N+1-n,M+1-m}|0>,
\end{eqnarray}
where $n$ and $m$ are the cell indexes $n\in[1,N]$ and $m\in[1,M]$.
Thus, one may say that the symmetry of the spectrum about $k^0_x$ is the 
consequence of the  inversion symmetry in the direct space of the hexagonal lattice.

Finally, we remark  that the symmetry $E_{k^0_x+k_x}= E_{k^0_x-k_x}$ shown by Fig.2a 
in the absence of the electric field (but at non-zero magnetic field) can be proved in a 
similar way, 
looking for an operator $P$ that this time commutes with the Hamiltonian. In this case,
$P$ looks like in Eq.(A.6) but with the plus sign between the two terms. 
We also note that
at $\mathcal{E}=0$ all the atomic energies are equal $\epsilon^{a,b}_m=0$, fact that 
helps to attain the commutation relation $[H,P]_{-}=0$ (for which the relation 
$\epsilon^a_{M+1-m}=\epsilon^b_m$ is needed).
The resulting reflecting point $k^0_x$ is the same as in the previous case.


\begin{thebibliography}{30}
\bibitem{Note-phi}  
The Peierls phase is obtained by integrating the 
vector potential along the A-B bond, and the result depends on the
chosen position of the coordinates origin. The phase in Eq.(1)
corresponds to the origin specified in Fig.1. Of course, the energy spectrum 
is independent of the choice of the coordinate origin.
\bibitem{Note-critic}
It is difficult to estimate the critical $\beta$ in this case
because of the  additional level broadening  due to confinement.
\bibitem{Note-shape} During the increase of $\mathcal{E}$, the bulk state distorts itself: 
it is squeezed 
in the direction parallel
to the field and stretches perpendicularly. At the resonance, the state meets the edge and the 
'shortcut edge state' is created.
\end{thebibliography}
\end{document}